\documentstyle[prb,preprint,
aps,psfig]{revtex}
\begin{document}
\draft


\title{Density functional study of the 
adsorption of K on the Cu(111) surface}
\author{K. Doll}
\address{Institut f\"ur Mathematische Physik, TU Braunschweig,
Mendelssohnstra{\ss}e 3, D-38106 Braunschweig}
\maketitle

\begin{abstract}
The adsorption of potassium on the Cu(111) surface
 in a $(2\times 2)$ pattern has been
simulated with all-electron full-potential density functional calculations.
The top site is found to be the preferred adsorption site, with
the other highly symmetric adsorption sites being nearly degenerate.
The  bond length from potassium to the nearest copper atom
is computed to be 2.83 \AA. Population analysis and density of states
indicate that there is no evidence for covalent bonding so that
the binding mechanism appears to be a metallic bond.

\end{abstract}

\pacs{ }

\narrowtext
\section{Introduction}

The study of adsorbates on metal surfaces is of high importance for
scientists as well as for industry with potential applications, for example
 in catalysis. Alkali metals are relatively simple systems to consider
as adsorbates. Initially, it was assumed that they would always
occupy high-coordination sites.
It was therefore a big surprise when a first system was 
discovered where Cs, adsorbed on the Cu(111) surface,
occupied the top site and not
the threefold hollow site \cite{Lindgren1983}. 
Meanwhile, further systems of alkali metals
with top adsorption have been found --- typically heavier alkali atoms (from 
K on) on closed packed surfaces (for a review, see references 
\onlinecite{DiehlMcGrathlang,DiehlMcGrathkurz}). 
The situation is thus more complex than in the case of halogens 
where the adsorption site on closed-packed surfaces
is usually the threefold hollow site. 

Due to the increase in computational power, the theoretical study
with methods based on density functional theory (DFT)
of these adsorbate systems has become feasible in the last few years.
For example, 
systems such as Na and K on the Al(111) surface\cite{NeugebauerScheffler}
 or Na on the Cu(111) surface \cite{CarlssonHellsing}
have been studied recently. 
 
In this article, we study the adsorption of potassium on the Cu(111) surface
as a model system for alkali adsorption on metal surfaces. This is
an extension of earlier work of halogens on metallic surfaces
(Cl on Cu(111) \cite{DollHarrison2000} and Cl on Ag(111) 
\cite{DollHarrison2001}). For the system Cu(111) (2$\times$2)-K, adsorption
on the top site was observed experimentally \cite{Adler1993}.
The addressed questions are therefore whether
first principles simulation is able to reproduce the preferred adsorption 
site,
the magnitude of the
energy splitting of different highly symmetric adsorption sites,
the geometry,
the charge of the adsorbate, and the mechanism of the bond.

The applied methods are all-electron full-potential
calculations with a local Gaussian
basis set. Mainly gradient corrected density functional 
calculations were performed, the energetically most favorable structure 
was additionally optimized with the local density approximation,
for comparison.

\section{Computational Parameters}

A local basis set formalism was used where the basis functions are
Gaussian type orbitals centered at the atoms as implemented in the code
 CRYSTAL \cite{Manual}. For Cu, a $[6s5p2d]$ all-electron basis set as
used in previous studies on copper metal and chlorine adsorption on
copper metal was employed \cite{KCuF3,DollHarrison2000}. 
The K all-electron basis set
from work on alkali halides was chosen \cite{Prencipe}, with the two outermost
$sp$-exponents replaced by two $sp$-exponents optimized
for the free K atom (0.29 and 0.08 a.u.)
 and an additional $d$-exponent (0.50 a.u.), resulting in a $[5s4p1d]$ basis
set. Compared to the calculations on alkali halides, more diffuse 
basis functions are
necessary because the potassium atom is not fully ionized and
a description of its $4s$ and $4p$ states is required. 
A further enlargement of the potassium basis set is achieved indirectly
 because basis functions from neighboring centers are used to describe the
charge distribution.
\cite{Basissatzdetail} 
The bulk of the calculations was done with
the gradient corrected exchange and correlation functional of Perdew and Wang
(GGA) \cite{Perdewetal}. For the energetically most favorable site,
the geometry was re-optimized at the level of the local density approximation
(LDA) with Dirac-Slater exchange \cite{LDA}  and
the Perdew-Zunger correlation functional \cite{PerdewZunger}, for comparison.

An auxiliary basis set was used to fit the exchange-correlation potential.
For copper, it consisted of 12 even tempered (i.e. the ratio  from one
to the next higher exponent was kept fixed)
 $s$-exponents
and 12 $p$-exponents in the range from 0.1 to 2000, 8 $d$-exponents 
and 8 $f$-exponents in the range from 0.1 to 100.
For potassium, the same $s$- and $p$-exponents were chosen, and 5 
$d$-exponents in the range from 0.8 to 100.

The adsorption was modeled by using a three-layer copper slab with the
copper atoms arranged as in the face-centered cubic (fcc) lattice,
at the GGA-optimized bulk Cu
lattice constant\cite{DollHarrison2000} of 3.63 \AA.
Potassium was adsorbed on one side of this slab. This is presently the
computational limit for desktop workstations with this approach. This
is similar to  earlier studies \cite{DollHarrison2000,DollHarrison2001} 
where
three layers of metal and a single-sided adsorbate layer
were found to be sufficient for a theoretical description. 
A supercell approach with a 
(2$\times$2) adsorption pattern as in the experiment was used 
\cite{Adler1993}. This slab was not repeated in the third dimension so
that the model is truly two-dimensional.
The copper
atoms in the top layer were allowed to relax vertically, an additional
horizontal shift as observed in the adsorption of K on Ni(111)
 \cite{Fisheretal}
would, however,
 require too much numerical effort to be simulated. For comparison,
the vertical relaxation of the copper atoms was simulated
 in two different ways:
simulations were performed where only a uniform relaxation of the top
copper layer was possible, and simulations where a different vertical
relaxation for the copper atoms in the top layer was possible (i.e. 
substrate rumpling). 

\begin{figure}
\caption{The structures considered for K, 
adsorbed on the Cu(111) surface, at a coverage of one fourth of a monolayer,
$(2\times 2)$ unit cell. The copper atoms in the top layer 
are displayed by open circles. In the top adsorption site,
potassium occupies the sites above the copper atoms with number 1
(filled circles).
Alternatively considered adsorption sites were the threefold hollow 
sites above atoms 2,3,4 (fcc or hcp hollow, circles with horizontal lines) or 
the bridge site above atoms 3 and 4
(circles with horizontal and vertical lines). Note that fcc and hcp
hollow sites can not be distinguished in this figure. 
The positions of the second and third layer copper atoms underneath
the potassium layer are necessary to differentiate between fcc and hcp hollow
 sites.}
\label{geometryfigure}
\centerline
{\psfig
{figure=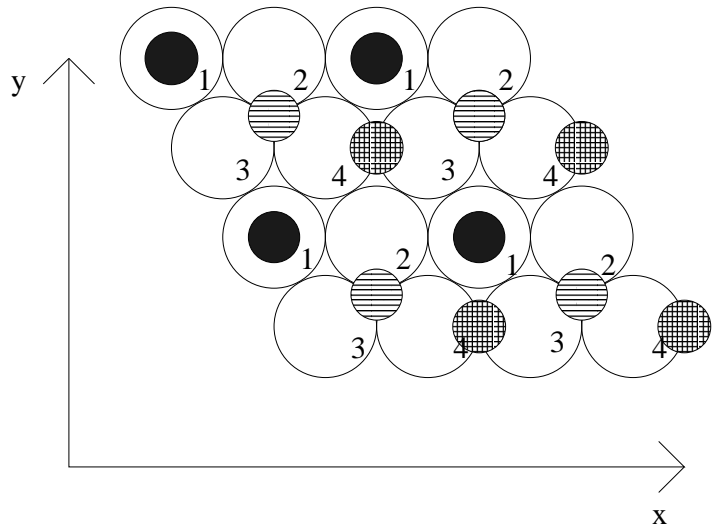,width=15cm,angle=270}}
\end{figure}

Four adsorption sites were considered (see figure
\ref{geometryfigure}): the top adsorption site with K sitting vertically
above a copper atom in the top layer, the bridge site with K sitting 
above the middle of two copper atoms in the top layer, and two
different threefold hollow sites
where the potassium atoms are placed vertically above 
a copper atom in the second (third) copper layer.
These threefold hollow sites can not be distinguished when only the adsorbate
and the positions of the atoms in the top copper layer are given, but instead
the position of the atoms in the second and third layer under the adsorbate
layer must be considered.
The first of the threefold hollow sites is
referred to as the hcp (hexagonal close-packed) hollow site because
the potassium atom is vertically above  a copper atom in the second copper 
layer.
The second of the threefold hollow sites is referred to as
fcc hollow site because the potassium atom is vertically above third
layer copper atoms. In the following, we will use the notion hcp (fcc) 
hollow site to distinguish between these two adsorption sites. This is closely
connected to the notion of hcp (versus fcc) lattices where there are
atoms two (versus three) layers vertically below atoms in a certain layer of
a solid.

With a Pack-Monkhorst net of shrinking factor 16,
the number of points in the reciprocal lattice was 30 for the fcc hollow site
and the
hcp hollow site, and 73 for the bridge and top site.
The Fermi function was smeared with a temperature of 0.01 $E_h$
(1$E_h$=27.2114 eV) to make the integration numerically more stable. 
For an extensive test of the various computational
parameters for metallic systems with a local basis set technique, 
see reference \onlinecite{KlausNicVic}.

\section{Summary of previous experimental and theoretical Results}

The adsorption of K on Cu(111) has been studied with surface-extended
X-ray absorption fine-structure (SEXAFS) measurements \cite{Adler1993}.
At a nominal coverage of 0.25$\pm$ 0.01 monolayer, a (2$\times$2) pattern
was observed.
The adsorption site was identified as the top site and a bond-length
from K to the nearest Cu atom of 3.05$\pm$0.02 \AA \ was deduced. 

Cluster models \cite{Padilla}
 with Cu$_7$ as the adsorbate gave in Hartree-Fock and
density functional calculations the top site as the preferential one.
However, the energy splitting was found to be relatively large
(e.g. $\sim$ 0.5 eV at the DFT level), the computed bond length was
in good agreement with the experimentally observed. The computed K charge
was relatively large with 0.7 - 0.8 $|e|$ at the DFT level. Similar results
for the energy splitting and bond length were then found when 
potentials were fitted to model the interaction of the
adsorbate with the infinite surface.

GGA calculations have been performed for the systems Cu(111) (2$\times$2)-Na
and Cu(111) (3/2$\times$3/2)-Na\cite{CarlssonHellsing}. 
It was found that hollow sites were preferred. 
Substrate rumpling was found to be
strongest for the top site; in general copper atoms below sodium atoms moved
into the substrate, whereas copper atoms without sodium in the vicinity
moved outwards. The top site was the one which gained most
energy from  rumpling and thus the energy difference between the
different adsorption sites was reduced.

The adsorption of sodium and potassium on the Al(111) surface has been studied
theoretically with the LDA \cite{NeugebauerScheffler}
and it was found that, whereas sodium adsorbed on a 
substitutional site, the fcc hollow site and the top adsorption site
were degenerate for potassium for both structures considered
($\sqrt{3} \times \sqrt{3}R30^\circ$ and 2$\times$2). Again, 
rumpling was most important for the top site.

\section{Results for K adsorption}
\label{potassiumadsorptionsection}

In this section, the results of the calculations on the system 
Cu(111) (2$\times$2)-K
are presented and discussed. Firstly,
in tables \ref{KonCuGGA} and \ref{KonCuGGAnorumpling}, 
the adsorption geometry and energy are displayed. 
We notice that the adsorption at the top site is energetically
most favorable, i.e. the experimental situation is reproduced. 
For this site, the computed adsorption energy per potassium atom
is 47 $mE_h$ (1.28 eV).
However,
the other considered sites are only slightly higher in energy (by less than
1 $mE_h$). This is close to the limit of what can be
resolved and there are still parameters which might have an influence
on the prediction of the adsorption site
(e.g. numbers of layers in the slab model, 
a possible horizontal relaxation of the copper atoms, basis set deficiencies
due to the lack of availability of $f$-functions). 
The magnitude of the energy
 splitting is thus much lower as previously found for halogens 
\cite{DollHarrison2000,DollHarrison2001}, but in good agreement with
findings for Na on Cu(111) \cite{CarlssonHellsing} and Na or K on Al(111)
\cite{NeugebauerScheffler}.

The value for the adsorption energy
is in the range of that for potassium on 
graphite (0.48 to 0.89 eV, depending on the coverage and the
relaxation\cite{Ancilotto}), or
that for Na on Cu(111) which was computed to be 1.8 to 1.9 eV
\cite{CarlssonHellsing}. 

The binding energy of a free K layer 
(with a K-K nearest neighbor distance as in the adsorbate system, 5.13 \AA)
was computed for comparison and a value of 0.0229 $E_h$, i.e.
0.62 eV was obtained. This is in reasonable agreement with previous calculations
(0.56 eV \cite{Ancilotto}, LDA, K-K nearest neighbor distance 4.94 \AA;
or 0.83 eV \cite{Wimmer}, LDA, K-K nearest neighbor distance 4.52 \AA;
in the latter work the energy of the free atom was 
obtained from a calculation without spin-polarization).

We notice that the K-Cu bond length increases from
2.83 \AA \ (top site) $<$ 3.04 \AA \ 
(bridge site) $<$ 3.11 \AA \ (fcc, hcp hollow site). This is in agreement with
the argument that the bond should be the stronger and the 
bond length therefore the shorter, 
the lower the number of copper neighbors \cite{Pauling}. It is consistent
with results from  simulations for halogens on metal surfaces 
\cite{DollHarrison2000,DollHarrison2001}.
In addition, we note that those atoms in the Cu surface, which are closest
to the K atom, relax inwards. This relaxation is strongest for the 
structure with the lowest number of 
nearest neighbors underneath the K atom  --- in full agreement with the
findings in the case of Na on Cu(111) \cite{CarlssonHellsing}.
The order of magnitude (0.165 \AA \  inwards relaxation of copper atom 1 under
the potassium atom, relative to atoms 2,3,4) is in line with 
Na on Cu(111) (0.1 \AA)\cite{CarlssonHellsing} and with one experiment
for K on Ni (111) (0.12 \AA) \cite{Fisheretal}, less with the other
experiment for K on Ni(111) (0.01 $\pm$ 0.09 \AA) \cite{Davis1994}.

The computed
bond length is shorter  than measured in the experiment (3.05 $\pm$
0.02 \AA)\cite{Adler1993}.  
To estimate the errors due to the choice of the functional,
calculations with the LDA were performed for
the top site, for comparison. 
The LDA bond length (2.73 \AA) is slightly shorter as the GGA bond length.
If we assume
a copper radius of 3.63/2$\sqrt{2}$ \AA, we obtain an 
effective radius, at the GGA level, of 1.55 \AA \ 
which is in the range of the experimentally observed
value for top adsorption of potassium (1.77 \AA \ for K on Cu(111),
1.57 ... 1.77 \AA \ for K on Ni(111)) --- see reference
\onlinecite{DiehlMcGrathlang} and references therein.

In addition, calculations for the same adsorption sites were
performed when rumpling was not allowed, but instead only a uniform relaxation
of the copper atoms in the top Cu layer (table \ref{KonCuGGAnorumpling}).
In these simulations,
the fcc hollow site was found to be lowest in energy, with the hcp hollow site
and
the bridge site of the order of $\sim$ 0.2 $mE_h$ higher in energy, i.e.
nearly degenerate. The top site was  highest in energy, but only by less than
 $1 mE_h$, however.
The bond lengths are relatively similar to the case where rumpling was
allowed, the atoms in the top Cu layer relaxed inwards by $\sim$ 0.03 \AA.
We can thus conclude that rumpling is crucial to obtain the top site
as the preferred site --- the top site gains $>2$ $mE_h$ due to 
rumpling, whereas all the other sites gain less than 1 $mE_h$. 

In table \ref{Kpopulationtable}, Mulliken charges for the potassium
atom are given, projected on the different basis functions. The free
atom in its $^2S$ state has 7 $s$ and 12 $p$ electrons, obviously
$p_x$, $p_y$ and $p_z$ are degenerate. In the free and neutral
K layer, $p_x$, $p_y$ and $p_z$
population slightly increase whereas the $s$ population decreases. 
Finally, in the adsorbate systems,
$p_x$, $p_y$ and $p_z$ charges are slightly above 4, the number of 
$s$-electrons is $\sim$ 6.5, and the total charge is
$\sim$ +0.1 $|e|$.
Charges in $p_x$ and $p_y$ orbitals are identical except for the bridge
site - as in the case of Cl on Ag(111)\cite{DollHarrison2001},
the $p_x$ orbital which has more overlap
with the substrate layer (in our choice of geometry --- figure
\ref{geometryfigure})
is slightly less occupied than the $p_y$ orbital. However, the difference
is much smaller than in the case of Cl on Ag(111), and it is hardly
visible in the density of states (DOS) ---
see the discussion in the following paragraph.
In addition, the overlap population is very small: for K and next
neighbor Cu, it is 0.06 $|e|$, and it rapidly decreases for the
copper atoms further apart. 
All this indicates that there is no evidence for hybridisation and covalent
bonding.

\begin{figure}
\caption{Density of states, on K projected, top adsorption site. The displayed
structure is the broad K $4sp$ band. 
The density of states, projected on all K basis functions, 
is shown with a full line.
A projection onto the K $s$-basis functions only
 is 
shown with a dashed-dotted line. The Fermi energy
is indicated with a dotted line.}
\vspace{1cm}
\label{DOSfigure}
{\psfig
{figure=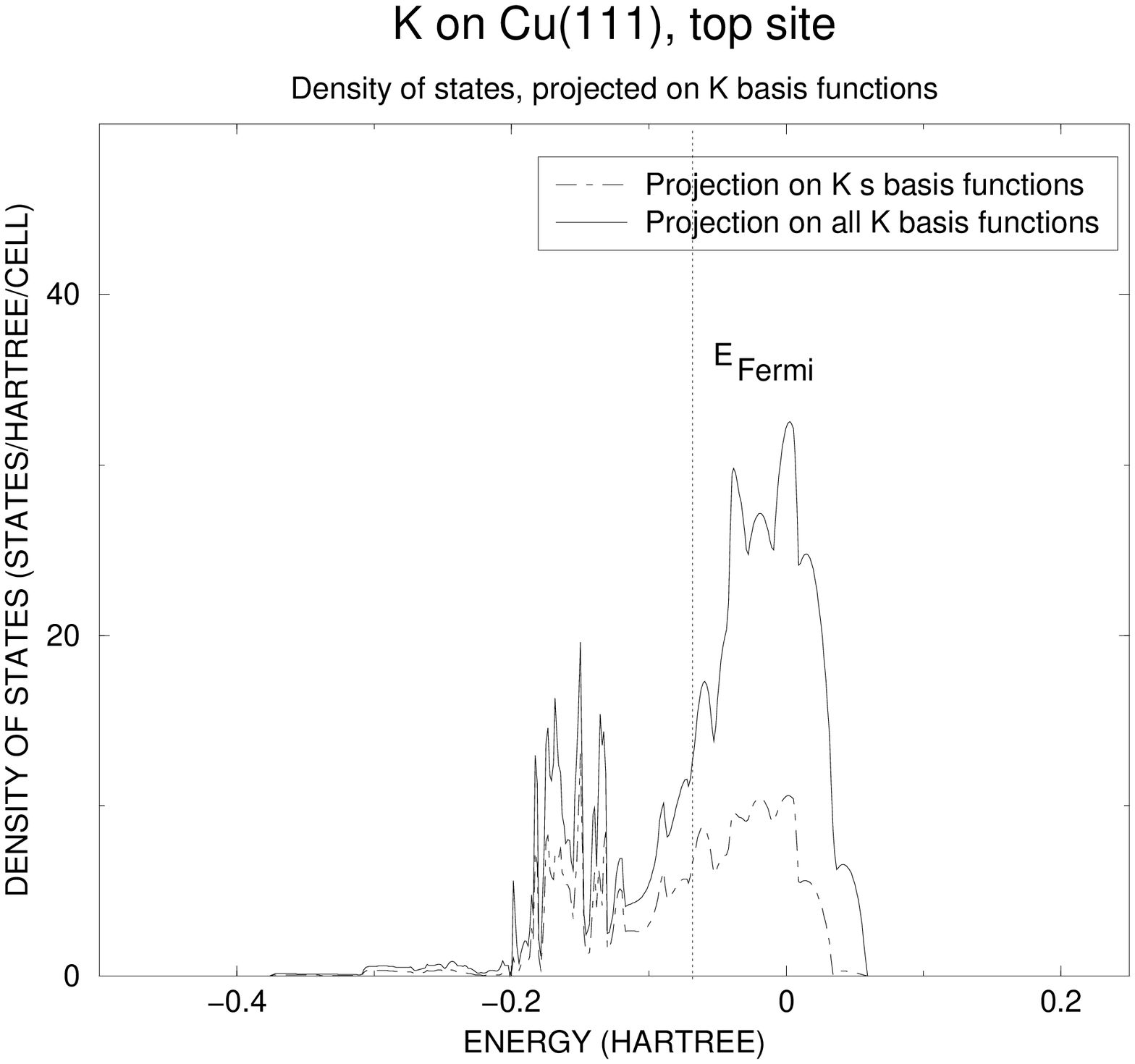,width=15cm,angle=270}}
\end{figure}

In figure \ref{DOSfigure}, the DOS, projected on the
potassium basis functions, is displayed for the top adsorption site. The Fermi
energy is in a regime with a relatively large contribution  from K 4$s$ and
4$p$ bands, so that the overlayer is clearly metallic. 
It is interesting that the projected DOS does not depend
on the adsorption site: it looks virtually
identical for top, hcp hollow, fcc hollow and bridge site.
This is different from the case of chlorine as an adsorbate, where 
the projected DOS clearly depended on the
adsorption site\cite{DollHarrison2001}. 
Another feature which was found for the bridge site in
the chlorine adsorption was the different projected DOS for
$p_x$ and $p_y$ orbitals: however, this is hardly visible for potassium
and the DOS for $p_x$ and $p_y$ is virtually identical. 

Also, the values
of the K $3s$ level
(-1.19 $E_h$ with respect to the Fermi level)
and $3p$ levels (-0.59 $E_h$ with respect to the Fermi level)
are practically independent of the adsorption
site. This is again in contrast to chlorine: for chlorine, the core eigenvalue
depended on the adsorption site, and it was correlated with the charge of
the chlorine atom \cite{DollHarrison2001}. For K, the variation of
the  Mulliken charges with adsorption site is much smaller (table
\ref{Kpopulationtable}) as for chlorine. This is consistent with
the finding that the core eigenvalue does not vary for the different
sites. The Mulliken charge for K at the top site is $\sim$ +0.12, i.e.
K is slightly positive charged. This is in reasonable agreement with
the findings for K on graphite, where, depending on the coverage and the
relaxation, 
charges in the range from 0.17 to 0.38 $|e|$, were obtained \cite{Ancilotto}.

\section{Summary}
In this article, all-electron full-potential density functional 
calculations were applied to compute properties of the system 
Cu(111) (2$\times$2)- K. The top site was found as the preferable
adsorption site, with the other considered sites being nearly degenerate.
Substrate rumpling was crucial to obtain this result. The bond length
was computed to be 2.83 \AA, the binding energy 1.28 eV. 
The potassium charge was computed to be +0.1 $|e|$.
Charge and projected density of states varied only very little for the
different adsorption sites, and similarly the position of the K $3s$ and $3p$
core eigenvalues did not vary. No evidence for a covalent contribution
to the binding was found, so that the mechanism should rather be a metallic
bond.


\onecolumn

\newpage
\begin{table}
\begin{center}
\caption{\label{KonCuGGA} Adsorption of K on the Cu(111) surface,
GGA results, if not stated otherwise. $d_1$ is the relaxation of those
atoms in the Cu top layer which are nearest
 neighbor to the adsorbed potassium, 
$d_2$ is the relaxation of those atoms which are not nearest neighbor to the
adsorbed potassium. The relaxations are given with respect to the bulk
value, a negative sign means a contraction relative to the bulk.
The bond length $d_{\rm {K-Cu \ nn}}$ 
is the distance from potassium to 
the nearest copper atom. 
The adsorption energy is the difference 
$E_{\rm {K \mbox{ }at \mbox{ } Cu(111)}}-{E_{\rm Cu(111)}-E_{\rm K}}$.
The numeration of the atoms is like in figure \ref{geometryfigure}.}
\begin{tabular}{ccccccc}
site &  nearest& relaxation  & 
atoms & relaxation  &
$d_{\rm {K-Cu \ nn}}$ & $E_{adsorption}$  \\
& atoms to K & $d_1$ [\AA] & not under K & $d_2$ [\AA] &  [\AA] & 
$\left[\frac{E_h}{K \ atom}\right]$ \\
hcp hollow$^a$ &
2,3,4 &   -0.045  &  1 &  +0.06   & 3.11 &  0.0464 \\
fcc hollow$^b$ &
2,3,4  &  -0.050  &  1 &  +0.07   & 3.11 &  0.0465 \\
bridge &
3,4    & -0.068  & 1,2 &  +0.03   & 3.04 &  0.0465 \\
top &
1    & -0.14   & 2,3,4 &  +0.025 & 2.83 &  0.0473 \\
\\
top (LDA) &  1  & -0.13 & 2,3,4 & +0.012  & 2.73 & 0.0660 \\
\\
\\
\end{tabular}
$^a$ note: Cu atom in third layer vertically under K atom \hfill \\
$^b$ note: Cu atom in second layer vertically under K atom \hfill
\end{center}
\end{table}

\newpage
\begin{table}
\begin{center}
\caption{\label{KonCuGGAnorumpling} Adsorption of K on the Cu(111) surface,
GGA results, without rumpling. The relaxation of the
atoms in the Cu top layer with respect to the bulk is given, the negative
sign means a contraction relative to the bulk, i.e. an inwards relaxation.
The bond length $d_{\rm {K-Cu \ nn}}$ is the distance from potassium to 
the nearest copper atom. 
The adsorption energy is the difference 
$E_{\rm {K \mbox{ }at \mbox{ } Cu(111)}}-{E_{\rm Cu(111)}-E_{\rm K}}$.}
\begin{tabular}{cccc}
site &   
 relaxation [\AA] &
$d_{\rm {K-Cu \ nn}}$ [\AA] & $E_{adsorption}$ 
$\left[\frac{E_h}{K \ atom}\right]$ \\
hcp hollow & -0.026 &   3.10 &  0.0457 \\
fcc hollow & -0.028 &   3.11 &  0.0459  \\
bridge &  -0.026 &  3.02 &   0.0457 \\
top &    -0.028   &   2.80 & 0.0451 \\
\\
free K layer & & & $E_{cohesive}$=0.0229 \\
\end{tabular}
\end{center}
\end{table}

\newpage
\begin{table}
\begin{center}
\caption{Orbital-projected charge of K on different adsorption sites.}
\label{Kpopulationtable}
\begin{tabular}{ccccccc}
site & \multicolumn{6}{c}{charge, in $|e|$} \\
& $s$ & $p_x$ & $p_y$ & $p_z$ & $d$ & total \\
 & \multicolumn{6}{c}{with rumpling}\\
top   & 6.484 & 4.155 & 4.155 & 4.072 & 0.012 & 18.879 \\
hcp hollow   & 6.481 & 4.149 & 4.149 & 4.072 & 0.012 & 18.863 \\
fcc hollow   & 6.481 & 4.150 & 4.150 & 4.072 & 0.012 & 18.865 \\
bridge & 6.481 & 4.147 & 4.155 & 4.072 & 0.013 & 18.867 \\
 & \multicolumn{6}{c}{without rumpling}\\
top   & 6.493 & 4.151 & 4.152 & 4.079 & 0.013 & 18.886 \\
hcp hollow   & 6.483 & 4.149 & 4.149 & 4.072 & 0.012 & 18.863 \\
fcc hollow   & 6.484 & 4.149 & 4.149 & 4.074 & 0.012 & 18.866 \\
bridge & 6.485 & 4.146 & 4.151 & 4.074 & 0.014 & 18.868\\
 & \multicolumn{6}{c}{free K layer}\\
       & 6.806 & 4.094 & 4.094 & 4.006 & 0.000 & 19.000 \\
 & \multicolumn{6}{c}{free K atom} \\
       & 7.000 & 4.000 & 4.000 & 4.000 & 0.000 & 19.000 \\
\end{tabular}
\end{center}
\end{table}

\newpage

\newpage

\begin{references}
\bibitem{Lindgren1983}S. \AA. Lindgren, L. Walld\'en, J. Rundgren,
P. Westrin and J. Neve, Phys. Rev. B {\bf 28}, 6707 (1983).
\bibitem{DiehlMcGrathlang}
R. D. Diehl and R. McGrath, Surf. Sci. Rep. {\bf 23}, 43 (1996).
\bibitem{DiehlMcGrathkurz}
R. D. Diehl and R. McGrath, J. Phys.: Condensed Matter
{\bf 9}, 951 (1997).
\bibitem{NeugebauerScheffler} J. Neugebauer and M. Scheffler,
Phys. Rev. B {\bf 46}, 16067 (1992).
\bibitem{CarlssonHellsing} J. M. Carlsson and B. Hellsing, Phys. Rev. B
{\bf 61}, 13973 (2000).
\bibitem{DollHarrison2000} K. Doll and N. M. Harrison, Chem. Phys. Lett.
{\bf 317}, 282 (2000).
\bibitem{DollHarrison2001} K. Doll and N. M. Harrison, Phys. Rev. B
{\bf 63}, 165410 (2001).
\bibitem{Adler1993} D. L. Adler, I. R. Collins, X. Liang, S. J. Murray,
G. S. Leatherman, K.-D. Tsuei, E. E. Chaban, S. Chandravarkar,
R. McGrath, R. D. Diehl and P. H. Citrin, Phys. Rev. B {\bf 48}, 17445
(1993).
\bibitem{Manual} V. R. Saunders, R. Dovesi, C. Roetti, M. Caus\`a, 
N. M. Harrison, R. Orlando, C. M. Zicovich-Wilson {\sc crystal 98} User's
Manual, Theoretical Chemistry Group, University of Torino (1998).
\bibitem{KCuF3} M. D. Towler, R. Dovesi, and V. R. Saunders, Phys. Rev. B
{\bf 52}, 10150 (1995).
\bibitem{Prencipe} M. Prencipe, A. Zupan, R. Dovesi, E. Apr\`a, and V. R.
Saunders, Phys. Rev. B {\bf 51}, 3391 (1995).
\bibitem{Basissatzdetail} This is different for the atom where a 
larger basis set is necessary to compute properties of the atom.
For calculations on the free K atom itself and for the calculations on
the free two-dimensional K layer, a $[6s5p1d]$ basis set was used 
(with outermost $sp$-exponents 0.39, 0.21, 0.03 instead of 0.29 and 0.08).
Calculations for the free atom were done with spin-polarized DFT, all
the other calculations were performed with non-spin-polarized DFT.
\bibitem{Perdewetal} J. P. Perdew, J. A. Chevary, S. H. Vosko,
K. A. Jackson, M. R. Pederson, D. J. Singh, and C. Fiolhais,
Phys. Rev. B {\bf 46}, 6671 (1992).
\bibitem{LDA} P. A. M. Dirac, Proc. Camb. Phil. Soc. {\bf 26}, 376 (1930);
J. C. Slater, Phys. Rev. {\bf 81}, 385 (1951); 
\bibitem{PerdewZunger} J. P. Perdew and A. Zunger, Phys. Rev. B {\bf 23},
5048 (1981).
\bibitem{Fisheretal} D. Fisher, S. Chandavarkar, I. R. Collins, R. D.
Diehl, P. Kaukasoina and M.  Lindroos, Phys. Rev. Lett. {\bf 68},
2786 (1992).
\bibitem{KlausNicVic} K. Doll, N. M. Harrison, and V. R. Saunders,
J. Phys.: Condensed Matter {\bf 11}, 5007  (1999).
\bibitem{Padilla}L. Padilla-Campos, A. Toro-Labb\'e, J. Maruani,
Surf. Sci. {\bf 385}, 24 (1997).
\bibitem{Ancilotto}F. Ancilotto and F. Toigo, Phys.Rev. B {\bf 47}, 
13713 (1993).
\bibitem{Wimmer} E. Wimmer, J. Phys. F {\bf 13}, 2313 (1983).
\bibitem{Pauling}L. Pauling, The nature of the chemical bond and the
structure of molecules and crystals, Cornell University Press (1960).
\bibitem{Davis1994} R. Davis, X.-M. Hu, D. P. Woodruff, K.-U. Weiss,
R. Dippel, K.-M. Schindler, Ph. Hofmann, V. Fritzsche and A. M. Bradshaw,
Surf. Sci. {\bf 307}, 632 (1994).
\end{references}
\end{document}